\documentclass[12pt]{article}
\usepackage{epsf}
\title{ {\bf
The inclusive $b\rightarrow s \gamma$ decay in the noncommutative standard 
model}}
\author{\vspace{1cm}\\
        {\bf E. O. Iltan}
        \thanks{E-mail address:
        eiltan@heraklit.physics.metu.edu.tr}
 \\
        Physics Department, Middle East Technical University \\
        Ankara, Turkey\\}

\date{}

\begin{document}
\setlength{\baselineskip}{24pt}
\maketitle
\setlength{\baselineskip}{7mm}
\begin{abstract}
We study the new structures appearing due to noncommutative effects in the 
inclusive decay $b\rightarrow s \gamma^*$, in the standard model. We 
present the corresponding coefficients which carries the space-space 
and space-time noncommutativity.   
\end{abstract} 
\thispagestyle{empty}
\newpage
\setcounter{page}{1}
\section{Introduction}
It is believed that the nature of the space-time changes at very short
distances of the order of the Planck length. The noncommutativity approach 
in the space-time is a possible candidate to describe the physics at the 
Planck scale. The noncommutative (NC) structure into space-time can be 
introduced by taking NC coordinates $\hat{x}_{\mu}$ which satisfy the
equation \cite{Synder}
\begin{eqnarray}
[\hat{x}_{\mu},\hat{x}_{\nu}]=i\,\theta_{\mu\nu} \, ,
\label{com1}
\end{eqnarray}
where $\theta_{\mu\nu}$ is a real and antisymmetric tensor with the 
dimensions of length-squared. Here $\theta_{\mu\nu}$ can be treated as 
a background field relative to which directions in space-time is 
distinguished.    

The noncommutative field theory is equivalent to the ordinary one except that
the usual product is replaced by the $*$ product 
\begin{eqnarray}
(f*g)(x)=e^{i\,\theta_{\mu\nu} \,\partial^y_{\mu}\,\partial^z_{\nu}} f(y)\,
g(z)|_{y=z=x}\,.
\label{product}
\end{eqnarray}
The commutation of the Hermitian operators $\hat{x}_{\mu}$ (see eq.
(\ref{com1})) holds with this new product, namely, 
\begin{eqnarray}
[\hat{x}_{\mu},\hat{x}_{\nu}]_*=i\,\theta_{\mu\nu} \,\, .
\label{com2}
\end{eqnarray}
The quantum field theory over noncommutative spaces \cite{Connes} has been 
reached a great interest in recent years with the re-motivation due to the 
string theory arguments \cite{Connes2,Witten}. Noncommutative field theories 
(NCFTs) are difficult to handle since they have non-local structure and 
the Lorentz symmetry is explicitly violated \cite{Mocioiu,Carlson1}. The 
violation of the Lorentz symmetry is due to the constants $\theta_{\mu\nu}$ 
in eq. (\ref{com1}). Since $\theta_{\mu\nu}$ is antisymmetric, the vectors 
$\theta_i=\epsilon_{ijk} \theta^{jk}$ and $\theta_{0i}$ are constant 
three-vectors in preferred directions in a given Lorentz frame.  

NCFTs have been studied extensively in the 
literature. There has been a lot of work done on the renormalizability of 
NCFT's \cite{Gonzales}. The unitarity in noncommutative theories and the 
unitarity properties of spontaneously broken noncommutative gauge theories 
have been discussed in \cite{Gomis} and \cite{Hewett} respectively. Bounding
noncommutative QCD due to the Lorentz violation has been studied in
\cite{Carlson1} and it was concluded that the collider limits were not
competitive with low energy tests of Lorentz violation for bounding the 
scale of space-time noncommutativity. Furthermore, noncommutativity among
extra dimensions for QED have been examined in \cite{Carlson2}.    
The noncommutative quantum electrodynamics (NCQED) have been studied in
\cite{Hayakawa} and the explicit calculation of electric dipole effects and 
anomalous magnetic moments have been done in \cite{Riad}. In the case of
non-abelian case, the field theory is formulated on noncommutative spaces
as theories on commutative spaces, by expressing the noncommutativity using
the $*$ product as in eq.(\ref{product}) \cite{Madore}. The method proposed in    
\cite{Madore} has been applied to the full Standard Model (SM) in
\cite{Calmet} and recently  a unique model for strong and electroweak
interactions with their unification has been constructed in \cite{Xiao}.
In a recent work \cite{Behr}, the SM forbidden $Z\rightarrow \gamma\gamma$ 
and $Z\rightarrow gg$ decays in the NCSM has been studied.   

In our work, we study the possible structures appearing for the process
$b\rightarrow s\gamma^*$ in the NCSM up to the first order in $\theta$,
using the consistent formalism of NCSM \cite{Calmet}.
Here the additional vertices of quarks with the scalar particles 
(in our case the scalar particle is the unphysical Higgs boson $\phi^{\pm}$) 
is proportional to the parameter $\theta_{\mu\nu} p^{\mu} q^{\nu}$  
where $p$ ($q$) is quark $(\phi^{\pm})$ four momentum vector. However, for 
the vertices of quarks with the vector particles, here W boson or 
photon with four momentum $q$, there exist new factors 
$\theta_{\alpha\mu} p^{\alpha} \slash \!\!\!{q}$ and 
$\theta_{\alpha\mu} q^{\alpha} \slash \!\!\!{p}$ in 
addition to  $\theta_{\alpha\beta} p^{\alpha} q^{\beta} \gamma_{\mu}$. The 
similar behavior appears for $\phi\phi\gamma$ and $WW\gamma$ vertices. 
Furthermore there are quark-quark-$\phi-\gamma$ and quark-quark-$W-\gamma$ 
four point interactions which do not exist in the commutative SM. Therefore 
new structures appear, in addition to the ones which are based on the 
assumption that NC effects enters in to the expressions as an exponential 
factor $e^{-\frac{i}{2}\, \theta_{\mu\nu} p^{\mu} q^{\nu}}$, which is 
consistent in approximate phenomenology (see \cite{Hinchliffe} and 
references therein).
\section{The noncommutative effects on the $b\rightarrow s\gamma^*$ decay}
\begin{figure}[htb]
\vskip 1.0truein
\centering
\epsfxsize=5.8in
\leavevmode\epsffile{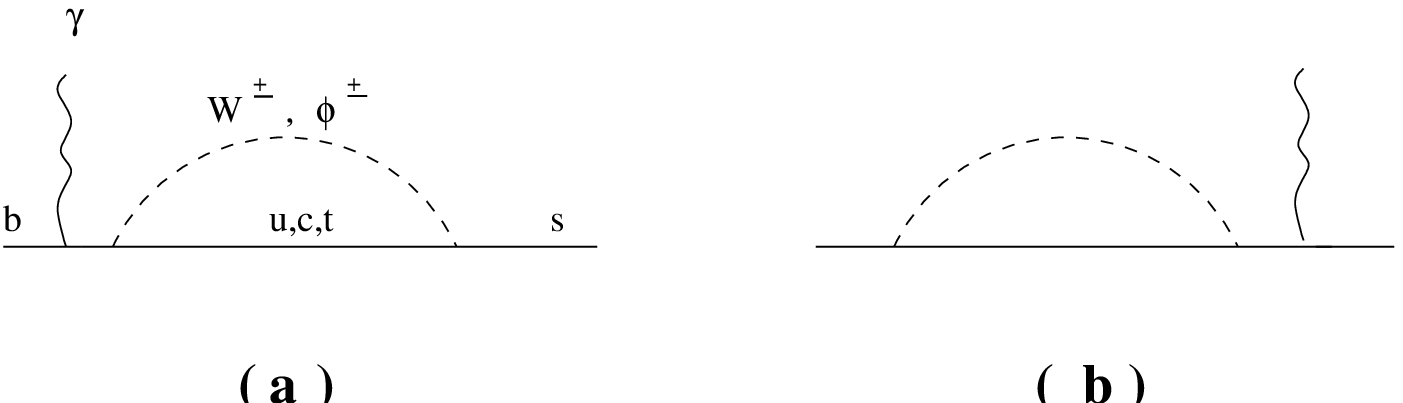}
\vskip 1.0truein
\caption[]{Self energy diagrams contribute to $b\rightarrow s \gamma^*$ in 
the NCSM. Wavy lines represent the elecromagnetic field and dashed lines the 
$W^{\pm}$ and $\phi^{\pm}$ fields.}
\label{Fig1}
\end{figure}

\begin{figure}[htb]
\vskip 1.0truein
\centering
\epsfxsize=5.8in
\leavevmode\epsffile{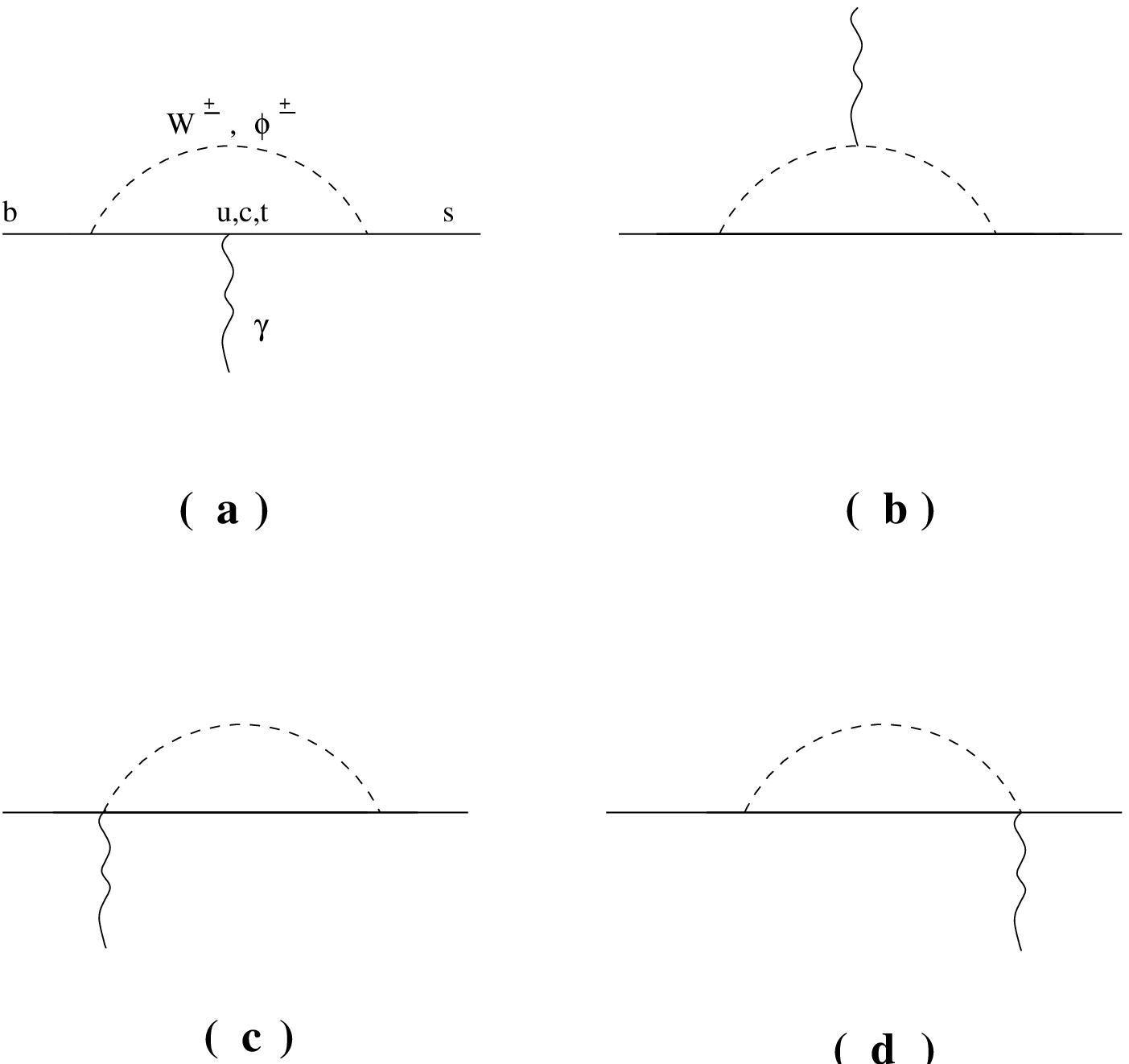}
\vskip 1.0truein
\caption[]{Vertex diagrams contribute to $b\rightarrow s \gamma^*$ in the 
NCSM.  Wavy lines represent the electromagnetic field and dashed lines the 
$W^{\pm}$ and $\phi^{\pm}$ fields.}
\label{Fig2}
\end{figure}

The inclusive $b\rightarrow s\gamma^*$ process appears at least in the loop 
level (see Fig. \ref{Fig1} and \ref{Fig2}). Now, we present the possible 
structures appearing for this process in the NCSM up to the first order 
in $\theta$:
\begin{eqnarray}
Q_1&=& \bar{s} (k_{\mu} \slash \!\!\!{k}-k^2 \gamma_{\mu} ) L b 
\nonumber \, , \\
Q_2&=& \frac{e}{8 \pi^2} m_b \bar{s} \sigma_{\mu\nu} \, k^{\nu} R b
\nonumber \, , \\
Q_3&=& m_b \bar{s} \widetilde{k}_{\mu} R b 
\nonumber \, , \\
Q_4&=& \bar{s} (k_{\mu} \slash \!\!\!{\widetilde{k}}-k^2 \theta_{\nu \mu}
\gamma_{\nu}) L b  \nonumber \, , \\
Q_5&=& \bar{s} (\widetilde{p}_{\mu} \slash \!\!\!{k}- \widetilde{p}.k
\gamma_{\mu}) L b  \nonumber \, , \\
Q_6&=& m_b \bar{s} (k_{\mu} \widetilde{p}.k-k^2 \widetilde{p}_{\mu}) R b 
\nonumber \, , \\
Q_7&=& m_b \bar{s}(\theta_{\nu\mu} \gamma_{\nu} \slash \!\!\!{k}- 
\slash \!\!\!{\widetilde{k}} \gamma_{\mu}) R b 
\nonumber \, , \\
Q_8&=& m_b \bar{s}(k_{\mu} \slash \!\!\!{k} \slash \!\!\!{\widetilde{k}}- 
k^2 \gamma_{\mu} \slash \!\!\!{\widetilde{k}}) R b 
\nonumber \, , \\
Q_9&=& m_b \bar{s} (k_{\mu} \slash \!\!\!{k} \slash \!\!\!{\widetilde{p}}- 
k^2 \gamma_{\mu} \slash \!\!\!{\widetilde{p}}) R b 
\nonumber \, , \\
Q_{10}&=& \epsilon_{\mu \alpha \beta \theta} k_{\beta} \widetilde{k}_{\alpha}  
\, \bar{s} \gamma_{\theta} L b 
\nonumber \, , \\
Q_{11}&=& \epsilon_{\mu \alpha \beta \theta} k_{\beta} \widetilde{p}_{\alpha}  
\bar{s} \gamma_{\theta} L b
\nonumber \, , \\
Q_{12}&=& (k^2 \epsilon_{\mu \alpha \beta \sigma}-
k_{\mu} k_{\theta} \epsilon_{\theta \alpha \beta \sigma})
\theta_{\beta\sigma} \, \bar{s}\gamma_{\alpha} L  b 
\nonumber \, , \\
Q_{13}&=& m_b k_{\theta} \epsilon_{\mu \theta \beta \sigma}
\theta_{\beta \sigma} \, \bar{s} R  b 
\label{structures}
\end{eqnarray}
where $L(R)=\frac{1-\gamma_5}{2}\, (\frac{1+\gamma_5}{2})$, $p$ ($k$) is the
four momentum vector of $b$ quark (photon $\gamma^*$) and 
$\widetilde{q}_{\mu}=\theta_{\mu\nu} q^{\nu}$. Here the first two structures 
exist in the commutative SM (CSM) and the others are due to the 
noncommutative effects. Notice that in the structures the $s$ quark mass is 
neglected.  For the real photon case, namely the $b\rightarrow s\gamma$ 
decay, the structures $Q_{2}$, $Q_{3}$, $Q_{5}$, $Q_{7}$, $Q_{10}$, $Q_{11}$ 
and $Q_{13}$ appear and the decay width for this process in the $b$-quark 
rest frame reads as 
\begin{eqnarray}
\Gamma &=&\Gamma_{CSM}+\Gamma_{New}
\label{DecayWidth1}
\end{eqnarray}
where
\begin{eqnarray}
\Gamma_{CSM}&=& \frac{G_F^2 \alpha_{em} m_b^5}{32\,\pi^4} 
|\sum_i A_2 (x_i)|^2 
\, ,\nonumber \\ 
\Gamma_{new}&=&\frac{G_F^2 \alpha_{em} m_b^3}{16\,\pi^4}
\Bigg(\epsilon_{\nu \alpha \beta \sigma} k_{\nu} p_{\alpha} 
\theta_{\beta \sigma} Re[\sum_i A_2^* (x_i)\,\sum_i A_7 (x_i)]-
(Im[\sum_i A_2 (x_i)\,\sum_i A_5^* (x_i)] 
\nonumber \\ &+& 
2 Im[\sum_i A_2 (x_i)\,\sum_i A_7^* (x_i)] + 
Re[\sum_i A_2^* (x_i)\,\sum_i A_{11} (x_i)])\, p.\widetilde{k} \, ,
\label{DecayWidth2}
\end{eqnarray}
with $i=u,c,t$ and $x_{i}=\frac{m_i^2}{m_W^2}$. Here 
$A_j (x_i)=V_{ib}\, V_{is}^* C^{NC}_j (x_i)$ and $C^{NC}_j (x_i)$ are the 
coefficients corresponding to the existing structures. The 
coefficient $C^{NC}_2 (x_i)$ is the well known Wilson coefficient 
$C_7 (x_i)$. 

It is obvious that the main contribution to the decay width comes from the
CSM  since the new part due to the NCSM is proportional to the extremely 
small parameter $\theta$. This new part is responsible for time-space and 
space-space noncommutativity. With the definitions 
$(\theta_T)_i=\theta_{0i}$ and $(\theta_S)_i= \epsilon_{ijk} \theta^{jk}$, 
$i,j,k=1,2,3$, in the $b$-quark rest frame, $\Gamma_{New}$ can be written 
as        
\begin{eqnarray}
\Gamma_{new}&=&\frac{G_F^2 \alpha_{em} m_b^3}{16\,\pi^4}
\Bigg(Re[\sum_i A_2^* (x_i)\,\sum_i A_7 (x_i)] \vec{k}.\vec{\theta_S}-
(Im[\sum_i A_2 (x_i)\,\sum_i A_5^* (x_i)]
\nonumber \\ &+&  
2 Im[\sum_i A_2 (x_i)\,\sum_i A_7^* (x_i)] + 
Re[\sum_i A_2^* (x_i)\,\sum_i A_{11} (x_i)])\, \vec{k}.\vec{\theta_T}\, .
\label{DecayWidth3}
\end{eqnarray}
This expression shows that the space-space noncommutativity is carried by
the coefficients $C^{NC}_2 (x_i)$ and $C^{NC}_7 (x_i)$. In the case of 
real coefficients $C^{NC}_{ 5,7,11} (x_i)$, $C^{NC}_2 (x_i)$ and 
$C^{NC}_{11} (x_i)$ play the main role for the time-space noncommutativity, 
since the imaginary parts of $A_j (x_i)$ are coming from the CKM matrix
elements, which are extremely small.      

In conclusion, the NC part of the decay width of the process under 
consideration is not easy to detect using present and even future sensibly 
arranged experiments. However, it brings a new source for the CP violating 
effects in addition to the complex CKM matrix elements in the SM, $V_{ub}$ 
in our case. With the consistent calculations of the coefficents of
this process and the precise experimental results of CP violating asymmetry, 
it would be possible to test the noncommutative effects and to predict the
noncommutative direction $\vec{\theta}_S$, if the the matrix 
$\theta_{\mu\nu}$ has constant components across the distances that are
large compared with the NC scale.   
\section{Acknowledgement}
This work was supported by Turkish Academy of Sciences (TUBA/GEBIP).

\newpage
\end{document}